\documentclass[aps,prl,twocolumn,showpacs,superscriptaddress,groupedaddress]{revtex4} 
\usepackage[a4paper, total={7in, 10in}]{geometry}
\usepackage{graphicx}
\usepackage{dcolumn}
\usepackage{bm}
\usepackage{amssymb}
\usepackage{amsmath}
\usepackage{empheq}
\usepackage[dvipsnames]{xcolor}
\usepackage{verbatim}
\usepackage{slashed}
\usepackage{MnSymbol}
\usepackage{float}
\usepackage[colorlinks=true,linkcolor=blue,urlcolor=black,bookmarksopen=true]{hyperref}

\newcommand{\wt}{\tilde{\omega}}

\begin{document}

\title{Equilibration of the Quark-Gluon Plasma at finite net-baryon density in QCD kinetic theory}
\author{Xiaojian Du}
\affiliation{Fakultät für Physik, Universität Bielefeld, D-33615 Bielefeld, Germany}
\author{Sören Schlichting}
\affiliation{Fakultät für Physik, Universität Bielefeld, D-33615 Bielefeld, Germany}
\date{\today}

\begin{abstract}
We explore the out-of-equilibrium dynamics of the Quark-Gluon Plasma at zero and finite net-baryon density based on an effective kinetic theory of Quantum Chromo Dynamics (QCD). By investigating the isotropization of the longitudinal pressure, we determine the relevant time and temperature scales for the onset of viscous hydrodynamics, and quantify the dependence on the chemical composition of the QGP. By extrapolating our results to realistic coupling strength, we discuss phenomenological consequences regarding the role of the pre-equilibrium phase at different collision energies.
\end{abstract}

\maketitle

\textit{Introduction}~---High-energy Heavy-Ion Collisions (HICs) at the Relativistic Heavy-Ion Collider (RHIC) and the Large Hadron Collider (LHC), provide a unique opportunity to explore the properties of strong interaction matter under extreme conditions. During the first ${\rm fm}/c$ of the space-time evolution of heavy-ion collisions, the non-equilibrium plasma of quarks and gluons created in the collision of heavy nuclei undergoes an out-of-equilibrium evolution \cite{Schlichting:2019abc,Berges:2020fwq}, before relativistic viscous hydrodynamics becomes applicable to describe the ensuing collective expansion of the near-equilibrium Quark-Gluon Plasma (QGP)~\cite{Gale:2013da,Heinz:2013th}. During the collective expansion, the hot and dense QGP is described in terms of macroscopic properties such as temperature $T$, chemical potentials $\mu$ and fluid velocity $u^{\mu}$ and cools down until on a time scale $\sim 10 {\rm fm}/c$, the temperatures and densities approach the deconfinement transition of QCD, eventually leading to the production of the final state hadrons measured in the RHIC and LHC experiments.

Even though the early pre-equilibrium phase represents a rather short period of time during the complex space-time evolution of a heavy-ion collision, it is imperative to understand the pre-equilibrium evolution of the QGP in order to establish a complete theoretical description of the reaction dynamics. To this end, significant progress has been made in recent years towards understanding the equilibration and onset of hydrodynamic behavior in high temperature QCD plasmas, from holographic studies of supersymmetric theories~\cite{Chesler:2008hg,Balasubramanian:2010ce,Heller:2011ju,Keegan:2015avk} to weakly coupled Yang-Mills~\cite{Kurkela:2014tea,Kurkela:2018wud,Heller:2016rtz,Almaalol:2020rnu} and QCD plasmas~\cite{Kurkela:2018oqw} as well as in various models~\cite{Blaizot:2020gql,Blaizot:2019scw,Blaizot:2017ucy,Kamata:2020mka,Strickland:2018ayk,Martinez:2012tu,Kurkela:2019set,Behtash:2020vqk}

Despite clear differences at the microscopic level, a common finding of different theoretical approaches is that macroscopic properties of the system, such as the non-equilibrium evolution of the energy-momentum tensor $T^{\mu\nu}$, can be accurately described by relativistic viscous hydrodynamics on a time scale $\tau_{\rm hydro}$, long before the system approaches local thermal equilibrium on time scales $\tau_{\rm eq} \gg \tau_{\rm hydro}$. Strikingly, it has been also observed, that already at (very) early times $\tau \ll \tau_{\rm Hydro}$, the non-equilibrium evolution of macroscopic quantities may become insensitive to the details of the initial conditions, and can be effectively described in terms of \emph{non-equilibrium attractors}~\cite{Heller:2015dha,Romatschke:2017vte}, which provides an accumulation point for the evolution of macroscopic quantities in out-of-equilibrium plasmas~\cite{Behtash:2019txb,Heller:2020anv}. While the existence of such attractors has been firmly established for different microscopic system undergoing a one-dimensional Bjorken expansion~\cite{Heller:2015dha,Denicol:2019lio,Romatschke:2017vte,Strickland:2017kux,Strickland:2018ayk,Giacalone:2019ldn,Almaalol:2020rnu,Heller:2020anv}, the study of non-equilibrium attractors remains an active research topic in theoretical physics and has lead to a number of interesting results~\cite{Romatschke:2017acs,Behtash:2020vqk,Denicol:2017lxn,Denicol:2019lio} aiming to understand and extend the range of applicability of effective macroscopic descriptions, such as relativistic viscous hydrodynamics. 

Beyond theoretical interest, the memory loss of macroscopic quantities also plays an important role in the phenomenological description of the pre-equilibrium stage of high-energy heavy-ion collisions~, e.g. to quantify the (approximate) amount of entropy production during the early pre-equilibrium stage which is directly connected to experimental measurements of the charged particle multiplicity in the final state~\cite{Giacalone:2019ldn}, or to describe the space-time evolution of the pre-equilibrium plasma macroscopically in KoMPoST~\cite{Kurkela:2018wud,Kurkela:2018vqr,NunesdaSilva:2020bfs,Gale:2020xlg}.

So far investigations of the pre-equilibrium dynamics of the QGP have focused primarily on kinetic equilibration of pure glue QCD~\cite{Kurkela:2014tea} with a recent extension to QCD~\cite{Kurkela:2018oqw} for charge neutral plasmas. By performing numerical simulations of the leading order effective kinetic theory of QCD~\cite{Arnold:2002zm} for gluons and light flavor $(u,d,s)$ quarks, we extend the theoretical treatment of the pre-equilibrium description QGP for the first time to finite net charge/net baryon density.
We investigate the existence of non-equilibrium attractors and describe their phenomenological consequences in the theoretical description of the pre-equilibrium stage. Within this letter, we focus on the aspects most relevant to heavy-ion phenomenology, and refer to our companion paper~\cite{Du:2020pre} for a detailed exposition of the theoretical framework and additional discussions.

\textit{Effective kinetic description of pre-equilibrium dynamics}~--- During the collision of heavy nuclei a fraction of the energy and valence charge is deposited in a primordial plasma, providing the initial conditions for the subsequent pre-equilibrium evolution of the QGP. In high-energy collisions the Color Glass Condensate (CGC) effective field theory of high-energy QCD~\cite{Gelis:2010nm,Iancu:2003xm}, provides a theoretical description of the gluon dominated initial state created at very early times~\cite{Kovner:1995ja,Krasnitz:1998ns,Krasnitz:2003jw,Blaizot:2010kh,Schenke:2012hg}. Despite recent efforts to generalize this framework to include quark production~\cite{Gelis:2015eua,Fujii:2006ab,Gelis:2005pb}, baryon stopping~\cite{McLerran:2018avb} and other corrections to the high-energy limit~\cite{Lushozi:2019duv,Kajantie:2019nse,Agostini:2019avp}, so far the chemical composition of the non-equilibrium plasma, as well as the structure of non-equilibrium initial state in heavy-ion collisions at lower beam energies or the forward rapidity regions of high energy collisions is currently not well understood. We will therefore consider a rather generic parametrization of the initial phase-space distributions of gluons, quarks and anti-quarks \footnote{We describe the pre-equilibrium evolution in terms of proper time $\tau=\sqrt{t^2-z^2}$ and rapidity $\eta=\text{atanh}(z/t)$ coordinates, along with the transverse $p_T=\sqrt{p_x^2+p_y^2}$ and longitudinal momenta $p_{\|}\equiv p^{\eta}/\tau=p_T\sinh(y_{p}-\eta)$ momenta in co-moving coordinates ($y_{p}=\text{atanh}(p_z/p)$). Since the pre-equilibrium stage only lasts for $\sim 1 {\rm fm}/c$, we follow the concept of \cite{Kurkela:2018vqr,Kurkela:2018wud} in neglecting the transverse expansion which becomes important at later times, and treating the pre-equilibrium QGP as locally homogenous in the transverse plane.}  
\begin{align}
&&f_{g}(\tau_0,p_{T},p_{\|})&=f^{0}_{g}
\frac{Q_0}{\sqrt{p_T^2+\xi_0^2p_{\|}^2}}e^{-\frac{2(p_T^2+\xi_0^2p_{\|}^2)}{3Q_0^2}}\;, \\
&&f_{q_{f}/\bar{q}_{f}}(\tau_0,p_{T},p_{\|})&=f^{0}_{q_{f}/\bar{q}_{f}} 
\frac{\sqrt{p_T^2+p_{\|}^2}}{\sqrt{p_T^2+\xi_0^2p_{\|}^2}}e^{-\frac{2(p_T^2+\xi_0^2p_{\|}^2)}{3Q_0^2}}\;.\nonumber
\end{align}
where the initial abundance of the $N_f=3$ light flavor quarks and anti-quarks is determined by the sum of the contributions from valence quark stopping ($f_{v}^{0}$) and quark/anti-quark pair production ($f_{0}^{s}$) as
\begin{align}
\label{eq:QuarkRatios}
f^{0}_{u}&=\frac{2n_p+n_n}{3(n_p+n_n)} f_{v}^{0}+ \frac{1}{2N_f} f_{s}^{0}\;, \qquad
f^{0}_{\bar{u}}=\frac{1}{2N_f} f_{s}^{0}\;, \nonumber \\
f^{0}_{d}&=\frac{n_p+2n_n}{3(n_p+n_n)} f_{v}^{0}+ \frac{1}{2N_f} f_{s}^{0}\;,  \qquad
f^{0}_{\bar{d}}=\frac{1}{2N_f} f_{s}^{0}\;, \nonumber \\
f^{0}_{s}&=\frac{1}{2N_f}f_{s}^{0}\;, \qquad \qquad \qquad 
f^{0}_{\bar{s}}=\frac{1}{2N_f}f_{s}^{0}\;, 
\end{align}
with $n_{p}/n_{n}=2/3$ denoting the proton/neutron fraction of the colliding nuclei. We will vary the parameters $\xi_0$ representing the initial momentum anisotropy, as well as $f^{0}_{g}$,$f^{0}_{s}$ and $f^{0}_{v}$ to investigate the sensitivity of our results to the initial conditions.  

Starting from the above initial conditions, we solve the QCD Boltzmann equation~\cite{Mueller:1999pi}
\begin{eqnarray}
\label{eq-bolzmannExp}
&&\left[\frac{\partial}{\partial \tau} - \frac{p_{\|}}{\tau}\frac{\partial}{\partial p_{\|}} \right] f_a(\tau, p_T,p_{\|})=\\
&& \qquad -C^{{2\leftrightarrow2}}_a[f](\tau, p_T,p_{\|})-C^{{1\leftrightarrow2}}_a[f](\tau, p_T,p_{\|})\;, \nonumber 
\end{eqnarray}
for a (longitudinally) boost invariant (transversely) homogeneous system undergoing a one dimensional Bjorken expansion. We include all leading order elastic $(C^{{2\leftrightarrow2}})$ and inelastic $(C^{{1\leftrightarrow2}})$ interactions, and following previous works~\cite{Kurkela:2018vqr,Kurkela:2018oqw} account for in-medium screening of elastic interactions~\cite{Arnold:2002zm} and the Landau-Pomeranchuk-Migdal~\cite{Landau:1953gr,Landau:1953um,Migdal:1955nv} suppression of inelastic rates, as described in detail in our companion paper~\cite{Du:2020pre}. 

\textit{Non-equilibrium evolution of macroscopic properties}~---
In order to connect the non-equilibrium initial state, with the subsequent hydrodynamic evolution we will focus on the pre-equilibrium evolution of the macroscopic properties of the QGP. Specifically, we will investigate the evolution of the energy momentum tensor $T^{\mu\nu}$ and conserved currents $J^{\mu}_{f}$, determined as
\begin{eqnarray}
\label{eq-energydensity}
\nonumber
T^{\mu\nu}&=&\int \frac{d^3p}{(2\pi)^3} \frac{p^{\mu}p^{\nu}}{p}\left[\nu_g f_{g}(\vec{p})+\nu_q \sum_f\left(f_{q_{f}}(\vec{p})+f_{\bar{q}_{f}}(\vec{p})\right)\right]\;,\\
\Delta J^{\mu}_{f}&=&\int \frac{d^3p}{(2\pi)^3} \frac{p^{\mu}}{p} \left[\nu_q\left(f_{q_{f}}(\vec{p})-f_{\bar{q}_{f}}(\vec{p})\right)\right]\;.
\end{eqnarray}
Based on the symmetries of the one-dimensional Bjorken expansion, one finds that in Milne $(\tau,x,y,\eta)$ coordinates $T^{\mu\nu}=\text{diag}(e,p_T,p_T,\frac{p_L}{\tau^2})$, $\Delta J^{\nu}_{f}=( \Delta n_{f},0,0,0)$ in the local rest-frame $u^{\mu}=(1,0,0,0)$, and the conservation laws take the form
\begin{eqnarray}
\label{eq-hydrodynamics-e}
\partial_{\tau}e+\frac{e+p_L}{\tau}=0\;, \qquad 
\label{eq-hydrodynamics-n}
\partial_{\tau}\Delta n_{f}+\frac{\Delta n_f}{\tau}=0,
\end{eqnarray}
such that irrespective of the underlying microscopic dynamics, the charge density per unit rapidity $\tau \Delta n_{f}=(\tau \Delta n_{f})_0=(\tau \Delta n_{f})_{\rm eq}$ remains constant throughout the evolution. Conversely, the evolution of the energy density $e$ is affected by the longitudinal pressure $p_L$. Due to the rapid longitudinal expansion, the longitudinal pressure $p_L \ll e$ is initially  much smaller than the transverse pressure $p_T \simeq e/2$, resulting in a constant ratio of energy per baryon $e/\Delta n_{B}$ and a constant energy density per unit rapidity $(e \tau)_0$ at early times. 

\begin{figure}[t]
	\begin{minipage}[b]{0.98\linewidth}
		\centering
		\includegraphics[width=1.00\textwidth]{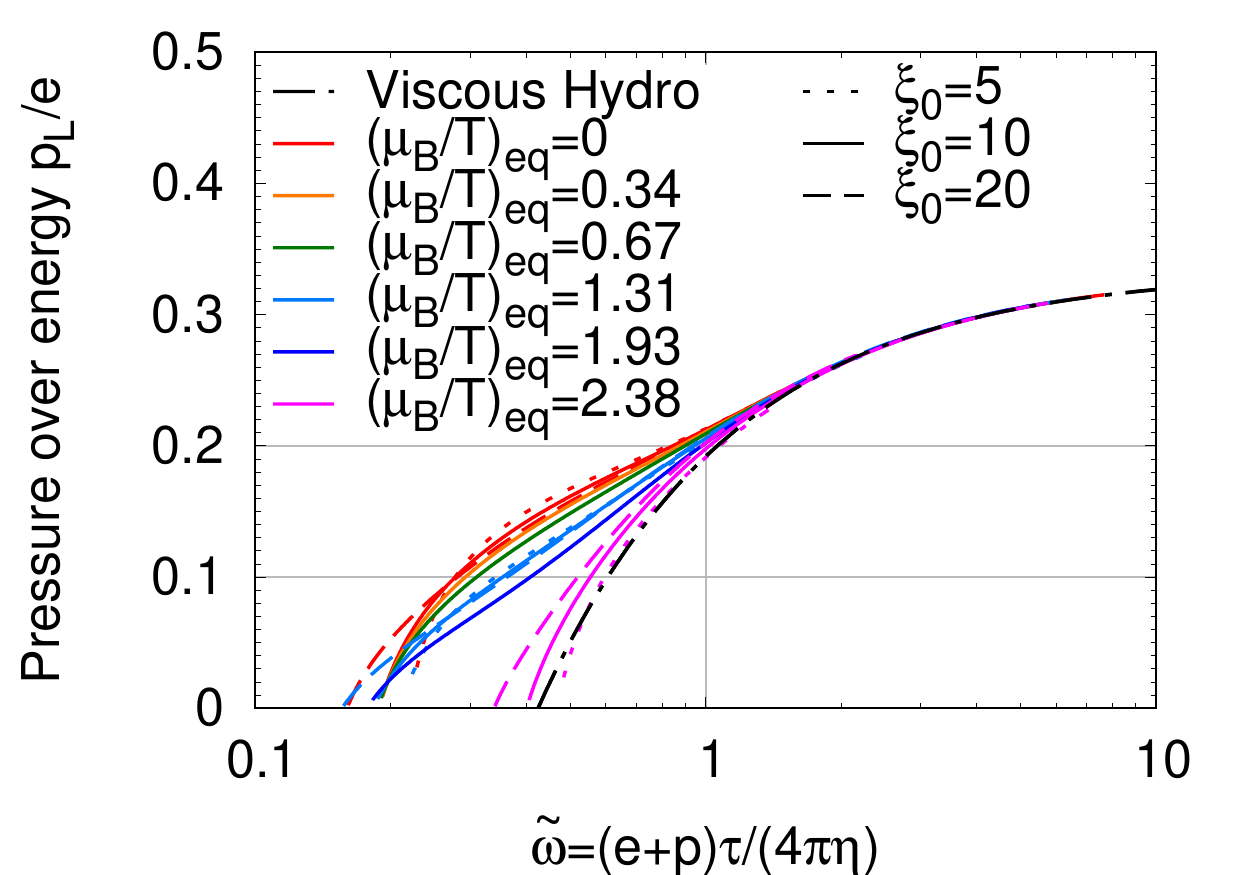}
	\end{minipage}
	\begin{minipage}[b]{0.98\linewidth}
		\centering
		\includegraphics[width=1.00\textwidth]{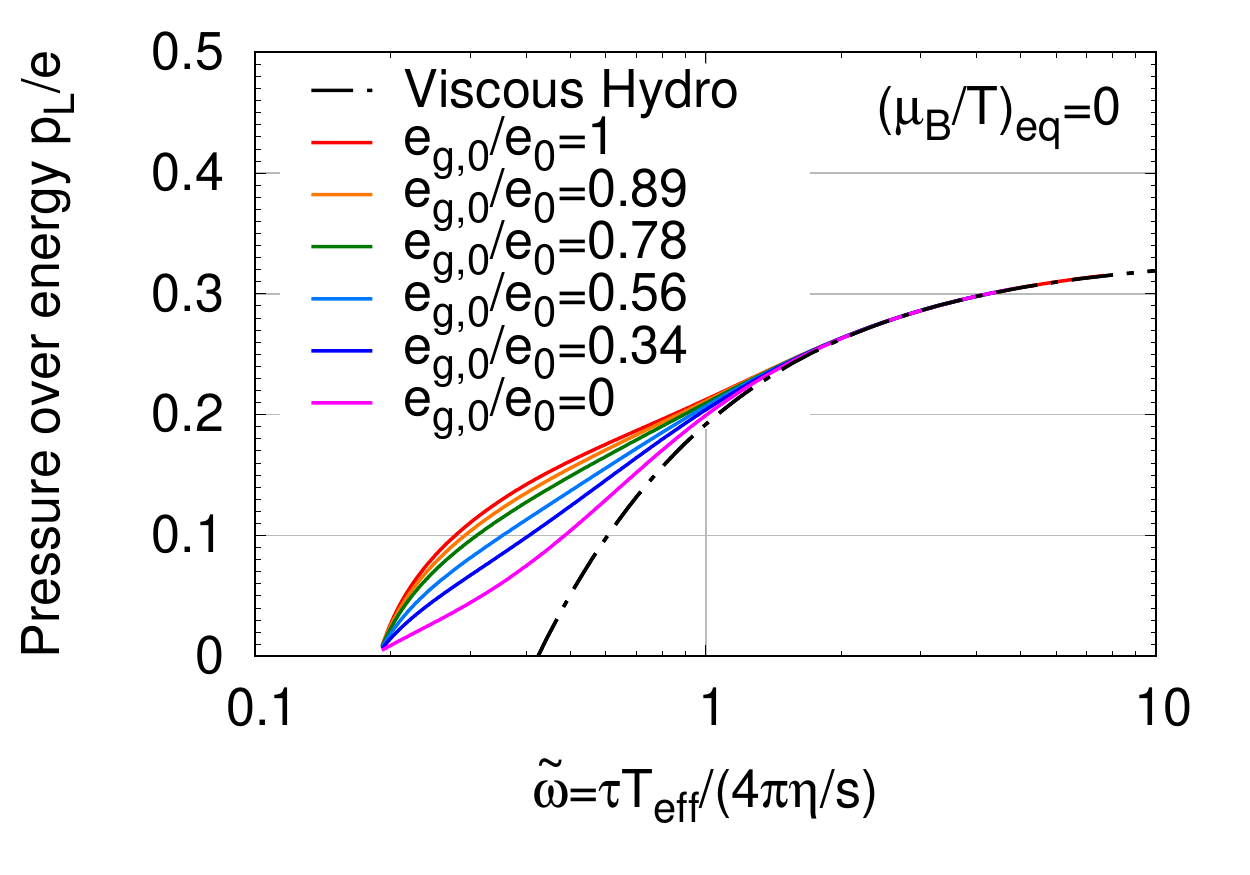}
	\end{minipage}
	\caption{Non-equilibrium evolution of the longitudinal pressure over energy ratio $\frac{p_L}{e}$ for different net-baryon density (top) and different initial quark to gluon ratios at zero density (bottom). Dotted and dashed curves in the top panel show results for different initial anisotropies $\xi_0$.}
	\label{fig-EXP-PLE}
\end{figure}

Over the course of the non-equilibrium evolution, the longitudinal expansion slows down, while kinetic interactions lead to a continuous increase of the longitudinal pressure $p_L$ (c.f. Fig.~\ref{fig-EXP-PLE}). Eventually, the residual deviations from equilibrium can be captured by hydrodynamic constitutive relations, where to first order in the gradient expansion the longitudinal pressure is determined by
\begin{eqnarray}
\label{eq-isotro-indicator}
\frac{p_L}{e}=\frac{1}{3}-\frac{16}{9}\frac{\eta}{(e+p)\tau}
\end{eqnarray}
with shear-viscosity $\eta$ and thermodynamic pressure $p=e/3$ for a conformal plasma. Ultimately, the QGP approaches an isotropic equilibrium state ($p_L=p_T=e/3$) where energy and charge densities are described by temperature $T$ and chemical potentials $\mu_{f}$ as
\begin{eqnarray}
\label{eq:EqTD}
e_{\rm eq}&=&\left[\nu_g\frac{\pi^2}{30}+\nu_q\frac{\pi^2}{120}\sum_f\left(7+\frac{30}{\pi^2}\frac{\mu_{f}^2}{T^2}+\frac{15}{\pi^4} \frac{\mu_{f}^4}{T^4}\right)\right]T^4\;, \nonumber \\
\Delta n_{f,eq}&=&\frac{\nu_q}{6}\left[\frac{\mu_{f}}{T}+\frac{1}{\pi^2}\frac{\mu_{f}^3}{T^3}\right]T^3
\end{eqnarray}
for an ultra-relativistic plasma of quarks and gluons, such that $\tau^{4/3}e$ and $\mu_{f}/T$ approach constant values $(\tau^{4/3}e)_{\rm eq}$, $(\mu_{f}/T)_{\rm eq}$ at late times $\tau \gg \eta/(e+p)$ when the QGP approaches local thermal equilibrium.

\begin{center}
\begin{table}
\begin{tabular}{||c|c|c|c|c|c|c|c|c||}
\hline
$Q_0 \tau_0$ & $f_g^0$ & $f_{val}^0$ & $f_{split}^0$ &
$\xi_0$ & 
$(\frac{\mu_B}{T})_{\rm eq}$ &
$\left(\frac{\eta T_{\rm eff}}{e+p}\right)$ & $C_{\infty}$\\ 
\hline \hline
9.9 & 1.068 & 0 & 0 & 
10 & 0 & 1.00 & 0.91\\
\hline
9.9 & 1.068 & 0 & 0 & 
5 & 0 & 1.00 & 0.93\\ 
\hline
9.9 & 1.068 & 0 & 0 & 
20 & 0 & 1.00 & 0.91\\ 
\hline
9.9 & 0.950 & 0 & 0.269 & 
10 & 0 & 1.00 & 0.92\\ 
\hline
9.9 & 0.833 & 0 & 0.534 & 
10 & 0 & 1.00 & 0.94\\ 
\hline
9.9 & 0.598 & 0 & 1.068 & 
10 & 0 & 1.00 & 0.96\\ 
\hline
9.9 & 0.363 & 0 & 1.602 & 
10 & 0 & 1.00 & 0.98\\ 
\hline
9.9 & 0 & 0 & 2.427 & 
10 & 0 & 1.00 & 1.01\\ 
\hline
9.9 & 0.950 & 0.269 & 0 & 
10 & 0.34 & 1.01 & 0.93\\ 
\hline
9.9 & 0.833 & 0.534 & 0 & 
10 & 0.67 & 1.01 & 0.94\\ 
\hline
9.9 & 0.598 & 1.068 & 0 & 
10 & 1.31 & 1.03 & 0.96\\ 
\hline
9.9 & 0.598 & 1.068 & 0 & 
5 & 1.31 & 1.03 & 0.98\\ 
\hline
9.9 & 0.598 & 1.068 & 0 & 
20 & 1.31 & 1.03 & 0.96\\ 
\hline
9.9 & 0.363 & 1.602 & 0 & 
10 & 1.93 & 1.06 & 0.98\\ 
\hline
22.5 & 0.363 & 1.602 & 0 & 
10 & 2.38 & 1.08 & 1.08\\ 
\hline
22.5 & 0.363 & 1.602 & 0 & 
5 & 2.38 & 1.08 & 1.11\\ 
\hline
22.5 & 0.363 & 1.602 & 0 & 
20 & 2.38 & 1.08 & 1.08\\ 
\hline

\end{tabular}
	\caption{Summary of initial conditions for QCD kinetic theory simulations for coupling strength $\lambda=g^2N_c=10$ along with the values of $(\mu_{B}/T)_{\rm eq}$, $(\eta T_{\rm eff})/(e+p)$ and $C_{\infty}$  extracted from fits to the asymptotic behavior of $\Delta n_{f}/e^{3/4}$ in Eq.~(\ref{eq:EqTD}), $p_L/e$ in Eq.~(\ref{eq-isotro-indicator}) and $(\tau^{4/3}e)/(\tau^{4/3}e)_{\rm eq}$ in Eq.~(\ref{eq-freestreaming}).}
	\label{tb-parameter}
\end{table}
\end{center}

We present our results for the non-equilibrium evolution of the long. pressure to energy density ratio in Fig.~\ref{fig-EXP-PLE}, where we show the time evolution of $p_L/e$ for different initial conditions at zero and non-zero net-baryon density summarized in Tab.~\ref{tb-parameter}. Starting from a highly anisotropic initial state at early times, the longitudinal pressure soon exhibits a rapid rise, followed by a slow approach towards isotropy $(p_L/e=1/3)$ at late times.

Strikingly, previous studies of the pressure isotropization in pure glue QCD~\cite{Kurkela:2014tea,Kurkela:2018vqr,Kurkela:2018wud} and QCD at zero density~\cite{Kurkela:2018oqw} have shown, that the non-equilibrium evolution of the energy momentum tensor at different coupling strength is governed by a universal scaling variable $\tilde{w}=\frac{\tau T}{4\pi\eta/s}$~\cite{Heller:2016rtz,Giacalone:2019ldn}, representing the ratio of the evolution time $\tau$, to the the equilibrium relaxation time $\tau_{R}=\frac {4\pi\eta/s}{T}$. By introducing an effective temperature $T_{\rm eff}=\left(\frac{30e(\tau)}{\pi^2\nu_{\rm eff}}\right)^{\frac{1}{4}}$ with $\nu_{\rm eff}=\nu_g+\frac{7}{4}\nu_qN_f$ characterizing the energy density of the non-equilibrium QGP, the definition of the scaling variable can be generalized to finite density systems as
\begin{eqnarray}
\label{eq-wtilde}
\tilde{\omega}=\frac{(e+p)\tau}{4\pi\eta}=\left(\frac{e+p}{\eta T_{\rm eff}}\right)\frac{\tau T_{\rm eff}}{4\pi}
\end{eqnarray}
such that the hydrodynamic evolution of the pressure in Eq.~(\ref{eq-isotro-indicator}) remains a universal function of the scaling variable $\frac{p_L}{e}=\frac{1}{3}-\frac{4}{9\pi\tilde{\omega}}$, irrespective of the net-baryon density. Indeed one observes from Fig.~\ref{fig-EXP-PLE} that eventually all curves converge towards the same viscous hydrodynamic behavior, which provides an accurate description of the pressure anisotropy for $\tilde{w} \gtrsim 1$. We note that albeit, the dimensionless transport coefficient $\left(\frac{\eta T_{\rm eff}}{e+p}\right)$, which at zero density reduces to $\eta/s$, exhibits a chemical potential dependence, we find that the variations of $\left(\frac{\eta T_{\rm eff}}{e+p}\right)$ are only at the $10\%$ level for the considered range of chemical potentials $(\mu_{B}/T)_{\rm eq} \lesssim 2.5$, as can be seen from Tab.~\ref{tb-parameter}, where we provide the results for $\left(\frac{\eta T_{\rm eff}}{e+p}\right)$ extracted from fits to the late time asymptotics of $p_L/e$ in Eq.~(\ref{eq-isotro-indicator}).

Despite their common late time behavior, some important differences emerge in the evolution of the pressure anisotropy at intermediate times ($\tilde{w}\lesssim 1$). By comparing the solid and dashed curves in the top panel of Fig.~\ref{fig-EXP-PLE}, we find that variations of the initial momentum anisotropy $\xi_0$ do not significantly affect the evolution of the pressure beyond very early times, which is in line with detailed earlier investigation in pure glue QCD~\cite{Almaalol:2020rnu}. Conversely, changes in the chemical composition of the primordial plasma do have a noticeable effect on the isotropization of the pressure at intermediate times, resulting in a moderate dependence on the net-baryon density $\mu_{B}/T$ shown in the top panel of Fig.~\ref{fig-EXP-PLE}. Generally, the isotropization of the pressure proceeds more and more slowly the larger the quark fraction, which is in line with early theoretical expectations~\cite{Biro:1981zi}, attributing the slower equilibration of the quark sector to the differences in color factors in elastic and inelastic scattering processes. Even in zero density systems, shown in the bottom panel of Fig.~\ref{fig-EXP-PLE}, the initial quark/gluon fraction affects the evolution of the pressure at intermediate times $\tilde{w} \lesssim 1$, challenging the existence of a universal non-equilibrium attractor, where the evolution of the pressure anisotropy becomes insensitive to the details of the initial conditions before the onset of hydrodynamic behavior~\cite{Romatschke:2017vte,Strickland:2018ayk,Kurkela:2018vqr,Kurkela:2019set,Almaalol:2020rnu}. Nevertheless, the variations of $p_L/e$ are still rather moderate, and it is equally important to point out that by the time $\tilde{w} \sim 1$ where hydrodynamics becomes applicable, differences in the initial chemical composition no longer effect the pressure evolution.

\textit{Connecting the initial state to hydrodynamics} ---
Now that we have established the evolution of the pressure anisotropy during the early pre-equilibrium phase, we can determine how the initial energy density $(e\tau)_0$ and net-charge density $(\Delta n_{f} \tau)_0$ affect the initial conditions for the subsequent hydrodynamic evolution. By following the arguments of~\cite{Giacalone:2019ldn}, the evolution of the conserved quantities $e$ and $\Delta n_{f}$ during the pre-equilibrium phase can be compactly expressed as
\begin{eqnarray}
\label{eq-exp-nonlinear}
\nonumber
\left(\tau^{\frac{4}{3}}e\right)(\tilde{\omega})
&=& \left(4\pi \frac{\eta T_{\rm eff}}{e+p}\right)^{\frac{4}{9}} \left(\frac{\pi^2\nu_{\rm eff}}{30}\right)^{\frac{1}{9}} (e\tau)_0^{\frac{8}{9}} C_{\infty}\mathcal{E}(\tilde{\omega})\;,\\
\left(\tau\Delta n_f\right)(\tilde{\omega})
&=&\left(\tau\Delta n_f\right)_{0}\;,
\end{eqnarray}
where as explained in the supplemental material, the function $\mathcal{E}(\tilde{\omega})=\frac{\tau^{4/3}e(\tau)}{(\tau^{4/3}e)_{\rm eq}}$ describes the non-trivial evolution of the energy density due to work performed against the longitudinal expansion \cite{Bjorken:1982qr,Gyulassy:1983ub,Giacalone:2019ldn}. Specifically, at early and late times $\mathcal{E}(\tilde{\omega})$ has the asymptotic behavior
\begin{eqnarray}
\label{eq-freestreaming}
\nonumber
&&\mathcal{E}(\tilde{\omega}\gg1)\simeq 1-\frac{2}{3\pi\tilde{\omega}}  \qquad (\text{visc. hydro})\\
&&\mathcal{E}(\tilde{\omega}\ll1)\simeq C_{\infty}^{-1}{\tilde{\omega}}^{\frac{4}{9}} \qquad (\text{free-streaming})
\end{eqnarray}
such that the constant $C_{\infty}$ in Eqns.~(\ref{eq-exp-nonlinear}) and (\ref{eq-freestreaming}) describes the efficiency with which the initial energy density per unit-rapidity $(e\tau)_0$ is converted into the thermal energy density at the onset of hydrodynamics.

\begin{figure}[t!]
	\begin{minipage}[b]{0.98\linewidth}
		\centering
		\includegraphics[width=1.00\textwidth]{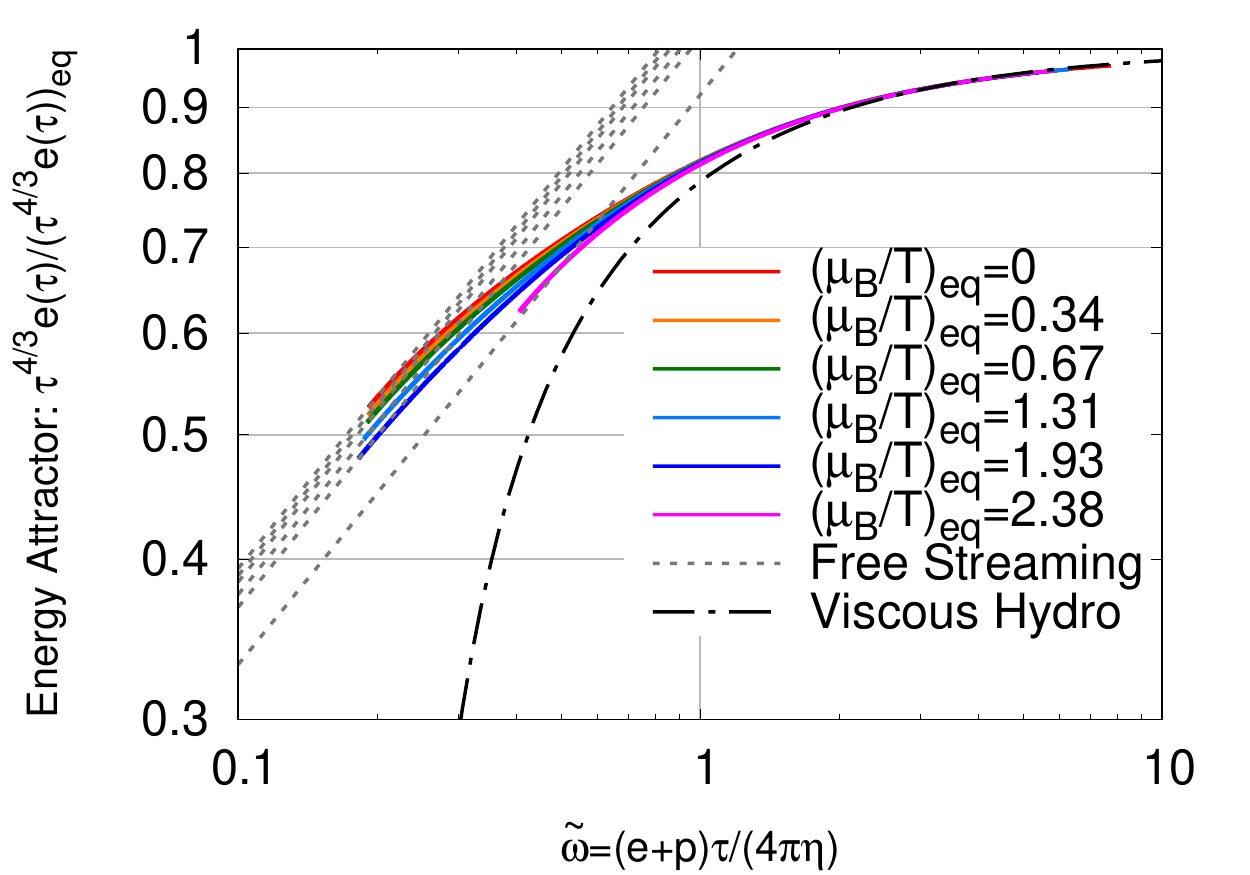}
	\end{minipage}
	\caption{Non-equilibrium evolution of the energy density for different net-baryon densities, characterized by the ratio of the chemical potentials to temperature $(\mu_{B}/T)_{\rm eq}$ at late times. Dashed and solid curves show a comparison to early and late time asymptotics in Eq.~(\ref{eq-freestreaming}).}
	\label{fig-EXP-ATT}
\end{figure}

Due to the different evolution of the pressure anisotropy, this conversion is different between zero and finite density systems, as can be seen in Fig.~\ref{fig-EXP-ATT}, where we present our results for the non-equilibrium evolution of the energy density of the QGP at different values of the baryon chemical potential $(\mu_{B}/T)_{\rm eq}$ at late times. Evidently the more baryon rich systems experience a smaller longitudinal pressure during the pre-equilibrium phase and therefore converts the initial energy density more efficiently into thermal energy.
By matching the results from our QCD kinetic theory simulations to the asymptotic behavior in Eq.~(\ref{eq-freestreaming}), we have extracted the values of $C_{\infty}$ to quantify this effect and provide the respective values for all simulations in Tab.~\ref{tb-parameter}. We find that for the considered range of parameters, the dependence on the chemical composition of the primordial plasma is typically on the $10-15\%$ level, indicating that despite the differences in the pressure evolution, the energy density at the beginning of the hydrodynamics phase can still be estimated to a rather good accuracy.

\textit{Phenomenological consequences} --- We conclude our analysis by studying the phenomenological consequences of our results, for the dynamical description of heavy-ion collision experiments at RHIC and LHC energies. Neglecting the entropy production during the late stage hydrodynamic expansion, we follow previous works~\cite{Giacalone:2019ldn,Jankowski:2020itt} matching the asymptotic entropy density
\begin{eqnarray}
(\tau s)_{\rm eq}= \frac{4}{3}\frac{(e\tau^{4/3})_{\rm eq}}{(\tau^{1/3}T)_{\rm eq}}-\left(\frac{\mu_{B}}{T}\right)_{\rm eq}\Delta n_{B}-\left(\frac{\mu_{Q}}{T}\right)_{\rm eq}\Delta n_{Q}\;,
\end{eqnarray}
to the experimentally measured charged particle multiplicities, according to $dN_{\rm ch}/d\eta =  \frac{N_{\rm ch}}{S} (\tau s)_{\rm eq} S_{\bot}$, and fix the final entropy to baryon number ratio $(\tau s)_{\rm eq}/(\tau \Delta n_{B})_{\rm eq}$, according to $\Delta n_{B}/s$=144, 285, 420 for 62.4\,GeV, 130\,GeV, 200\,GeV Au+Au collisions. We employ $S/N_{\rm ch}=8.36$ ~\cite{Giacalone:2019ldn,Hanus:2019fnc} and estimate the transverse area $S_{\bot}=\pi R_{A}^{2}$=128\,fm$^2$, 138\,fm$^2$, 95\,fm$^2$ for central Au+Au, Pb+Pb, Xe+Xe collisions with $dN_{\rm ch}/d\eta$=470, 590, 665 for 62.4\,GeV, 130\,GeV, 200\,GeV Au+Au collisions; $dN_{\rm ch}/d\eta$=1600, 1942 for 2.76\,TeV, 5.02\,TeV Pb+Pb collisions, $dN_{\rm ch}/d\eta$=1166 for 5.44\,TeV Xe+Xe collision~\cite{Arsene:2004fa,Back:2005hs,Aamodt:2010cz}. Since the overall normalization is fixed by this procedure, Eq.~(\ref{eq-exp-nonlinear}) then provides the evolution of the energy $e$ and charge densities $\Delta n_{f}$, from which we infer temperatures $T$ and baryon chemical potential $\mu_{B}$ based on the usual Landau matching procedure associating the conserved charges with their equilibrium values $e=e_{\rm eq}(T,\mu_{f})$ and $\Delta n_{f}=n_{\rm eq}(T,\mu_{f})$.

\begin{figure}[t!]
	\begin{minipage}[b]{0.98\linewidth}
		\centering
		\includegraphics[width=1.00\textwidth]{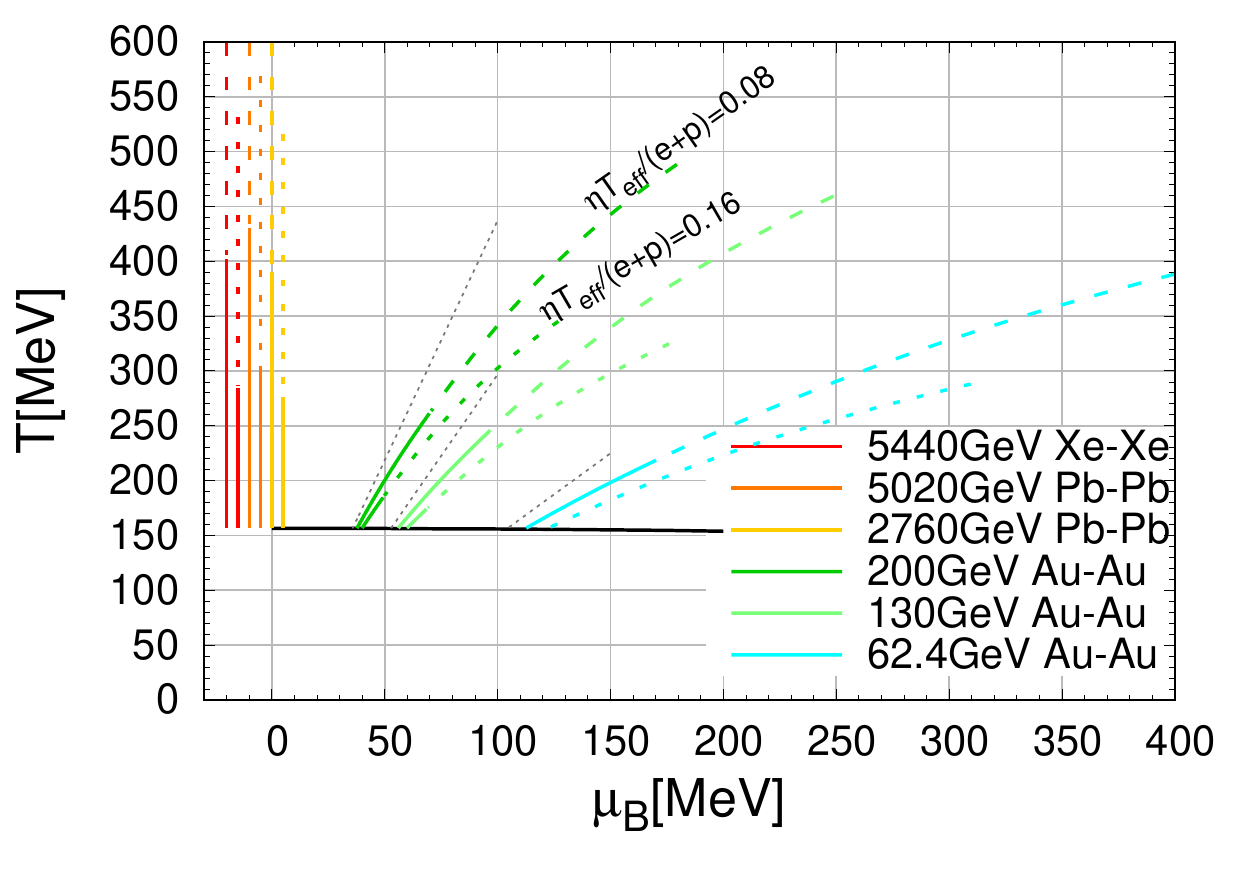}
	\end{minipage}
	\caption{ Non-equilibrium trajectories of temperature $T$ and baryon chemical potential $\mu_{B}$ in the QCD phase-diagram for $0-5\%$ most central heavy-ion collisions. Dashed/dotted lines show the pre-equilibrium evolution  $0.2<\tilde{w}<1$ for $\eta T_{\rm eff}/(e+p)$=0.08, 0.16, while solid lines correspond to the hydrodynamic evolution for $\tilde{w}>1$ which approaches the (perturbative) isentropes indicated by gray dashed lines. Black solid line indicates the QCD cross over temperature $T_{c}$=156.5\,MeV~\cite{Bazavov:2018mes}.}
	\label{fig-EXP-MUT}
\end{figure}

We illustrate the resulting non-equilibrium trajectories of the QGP in the QCD phase-diagram in Fig.~\ref{fig-EXP-MUT}, where dashed (dotted) lines represent the pre-hydrodynamic evolution of $(T,\mu_{B})$ for $0.2<\wt<1$, whereas solid lines show the hydrodynamic trajectories for $\wt>1$ for two different values of the transport coefficient $\eta T_{\rm eff}/(e+p)$=0.08 (0.16). Strikingly, one observes at first sight, that the pre-equilibrium plasma can exhibit much higher temperatures and chemical potentials as can be achieved during the subsequent hydrodynamic evolution. While in high-energy collisions of heavy nuclei, the trajectories run straight down along the vertical axis $(\mu_{B}/T\approx 0)$, the non-equilibrium trajectories at lower energies bend towards larger values of $\mu_{B}/T$ and deviate significantly from the (perturbative) isentropes indicated by gray dotted lines. 

Concerning the applicability of hydrodynamics, one finds that e.g. in high-energy Pb+Pb collisions, the hydrodynamic description becomes applicable on time scales 
\begin{eqnarray}
\nonumber
\tau &\simeq& 1.3~{\rm fm}/c \left(\frac{4\pi \eta/s}{2}\right)^{\frac{3}{2}} \left(\frac{dN_{ch}/d\eta}{1942}\right)^{-\frac{1}{2}} \left(\frac{S_{\bot}}{138 {\rm fm}^2}\right)^{\frac{1}{2}}
\end{eqnarray}
and temperatures 
\begin{eqnarray}
T &\simeq& 300 {\rm MeV}  \left(\frac{4\pi \eta/s}{2}\right)^{-\frac{1}{2}} \left(\frac{dN_{ch}/d\eta}{1942}\right)^{\frac{1}{2}} \left(\frac{S_{\bot}}{138 {\rm fm}^2}\right)^{-\frac{1}{2}}
\end{eqnarray}
well above the QCD cross-over temperature $T_{c}\simeq$156\,MeV~\cite{Bazavov:2018mes}, indicated by a black line in Fig.~\ref{fig-EXP-MUT}. However, at lower energies the non-equilibrium  trajectories extend almost all the way to the QCD phase boundary, indicating that the QGP created in low energy collisions may be significantly out of equilibrium for a substantial part of its lifetime. Even though the use of a perturbative description and the modeling of the space-time dynamics in terms of a one dimensional Bjorken expansion become increasingly questionable at lower energies, we still believe that Fig.~\ref{fig-EXP-MUT} points to one of several important challenges in describing the space-time evolution of heavy-ion collisions at low energies~\cite{Romatschke:2009im,Gale:2013da,Yan:2017ivm,Shen:2020mgh,Dore:2020jye}. 

\textit{ Conclusions \& Outlook}~---
We investigated the early time pre-equilibrium dynamics of the QGP at zero and finite densities of the conserved baryon and electric charge 
based on an effective kinetic description of high-energy QCD. We find that for a plasma undergoing a one dimensional Bjorken expansion, the non-equilibrium evolution of the energy momentum tensor can be well described by viscous hydrodynamics for time $\tau \gtrsim \frac{4\pi \eta}{e+p}$, which is in accordance with earlier findings at zero density~\cite{Kurkela:2018oqw}. However, the chemical composition of the QGP can have a significant impact on the evolution of the pressure at early times $\tau \lesssim \frac{4\pi \eta}{e+p}$, which is in sharp contrast to the results obtained for single component plasmas~\cite{Heller:2011ju,Keegan:2015avk,Kurkela:2014tea,Heller:2015dha,Kurkela:2018wud,Heller:2016rtz,Almaalol:2020rnu,Berges:2013eia,Berges:2013fga,Kamata:2020mka,Strickland:2018ayk,Denicol:2019lio,Romatschke:2017vte,Strickland:2017kux,Heller:2020anv}, where macroscopic properties such as $p_L/e$ exhibits a rapid memory loss at early times, resulting in a universal approach towards viscous hydrodynamics. Even though, the ensuing differences in the pressure only affect the late time evolution of the QGP at the $10-15\%$ level, it is clear that further progress in the theoretical understanding hinges to a considerable extent on an improved determination of the chemical composition of the primordial plasma, created immediately after the collision of heavy nuclei. 

We further discussed how the results presented in this letter can be used to connect the properties of the non-equilibrium initial state to the initial conditions for the subsequent hydrodynamic evolution. By extending previous works~\cite{Giacalone:2019ldn} to include the effects of finite net-baryon and electric charge density, our calculations thus provide a first important step towards including the evolution of all QCD conserved charges into dynamical descriptions of the pre-equilibrium stage of heavy-ion collisions~\cite{Kurkela:2018wud,Kurkela:2018vqr,Martinez:2019jbu}. Beyond such applications to bulk phenomenology, our QCD kinetic theory studies also provide the basis for future investigations of heavy flavor dynamics or the emission of electromagnetic probes, which could provide sensitive probes of the early time non-equilibrium dynamics~\cite{Churchill:2020uvk,Kasmaei:2018oag,Kasmaei:2019ofu}.

{\it Acknowledgement}~---
We thank Giuliano Giacalone and Aleksas Mazeliauskas for many insightful discussions and valuable contributions to the supplemental material. We further thank Aleksi Kurkela, Jean-Francois Paquet, Ismail Soudi and Derek Teaney for insightful discussions and collaboration on related projects. This work is supported by the Deutsche Forschungsgemeinschaft (DFG, German Research Foundation) – project number 315477589 – TRR 211. The authors also acknowledge computing time provided by the Paderborn Center for Parallel Computing (PC2) and the National Energy Research Scientific Computing Center, a DOE Office of Science User Facility supported by the Office of Science of the U.S. Department of Energy under Contract No. DE-AC02-05CH11231.

\section*{Supplemental material}
Below we provide a brief re-derivation of Eq.~(\ref{eq-exp-nonlinear}) which, as originally derived in Ref.~\cite{Giacalone:2019ldn}, relates the initial energy density to the initial conditions for the subsequent hydrodynamic evolution. Denoting $p_L/e=f(\wt)$ in the following with $\wt=\frac{T_{\rm eff}(\tau)\tau}{4\pi\eta/s}$ as in Eq.~(\ref{eq-wtilde}), the energy conservation equation can be re-cast into the form
\begin{eqnarray}
\tau \partial_{\tau} e = -e(1+f(\wt))\;,
\end{eqnarray}
Based on the knowledge of $p_L/e$, one can then directly integrate out the conservation equation. By changing variables from $\tau$ to $\wt$ according to
\begin{eqnarray}
\tau \partial_{\tau}=a(\wt)\wt\partial_{\wt}\;, \qquad 
a(\wt)=\frac{3}{4}-\frac{1}{4}f(\wt)
\end{eqnarray}
one obtains
\begin{eqnarray}
e(\wt_\tau)=e(\wt_0) \exp\left( - \int_{\wt_0}^{\wt_{\tau}} \frac{d\wt}{\wt} \frac{1+f(\wt)}{a(\wt)}\right).
\end{eqnarray}
Now in order to turn this into a more practical formula, it is useful to factor our the asymptotic behavior at early and late times, where obviously
\begin{eqnarray}
e(\wt\gg1) \propto \wt^{-2} 
\end{eqnarray}
\begin{eqnarray}
e(\wt\ll1) \propto \wt^{-\frac{4}{3}}
\end{eqnarray}
By considering the ratio $\frac{\wt_{\tau}^2 e(\wt_\tau)}{\wt_{0}^{\frac{4}{3}} e(\wt_0)}$, and making use of the relation between the energy density and the effective temperature $e(\wt)=\frac{\pi^2}{30}\nu_{\rm eff} T_{\rm eff}^4(\wt)$ to re-express additional powers of temperature in terms of additional powers of energy, one arrives at
\begin{eqnarray}
\label{eq:EnergyComplicated}
\tau^{\frac{4}{3}}e(\tau)&=&  
\left(4\pi \frac{\eta T_{\rm eff}}{e+p}\right)^{\frac{4}{9}} \left(\frac{\pi^2}{30}\nu_{\rm eff}\right)^{\frac{1}{9}} \left(e\tau\right)_{0}^{\frac{8}{9}} \\
 && \times \left[ \frac{\wt_{\tau}^{2}}{\wt_0^{\frac{4}{3}}}\exp\left( - \int_{\wt_0}^{\wt_{\tau}} \frac{d\wt}{\wt}  \nonumber \frac{1+f(\wt)}{a(\wt)}\right)\right]^{\frac{2}{3}}
\end{eqnarray}
where the expression in the square bracket is constructed in such a way that the limits $\wt_{\tau}\to\infty$ and $\wt_{0}\to0$ are finite, yielding 
\begin{eqnarray}
C_{\infty}=\lim_{\wt_{\tau}\to\infty, \wt_{0}\to0}\left[ \frac{\wt_{\tau}^{2}}{\wt_0^{\frac{4}{3}}}\exp\left( - \int_{\wt_0}^{\wt_{\tau}} \frac{d\wt}{\wt} \frac{1+f(\wt)}{a(\wt)}\right) \right]^{\frac{2}{3}}
\end{eqnarray}
By taking the limit $\wt_0 \to 0$, one can therefore re-express Eq.~(\ref{eq:EnergyComplicated}) as in~\cite{Giacalone:2019ldn}
\begin{eqnarray}
\tau^{\frac{4}{3}}e(\tau)= \left(4\pi\frac{ \eta T_{\rm eff}}{e+p}\right)^{\frac{4}{9}} \left(\frac{\pi^2}{30}\nu_{\rm eff}\right)^{\frac{1}{9}} \left(e\tau\right)_{0}^{\frac{8}{9}} C_{\infty} \mathcal{E}(\wt_{\tau})
\end{eqnarray}
with $\mathcal{E}(\wt_{\tau})= \frac{\tau^{\frac{4}{3}}e(\tau)}{\tau^{\frac{4}{3}}(e)_{\rm eq}}$ formally given by
\begin{eqnarray}
\mathcal{E}(\wt_{\tau})=\lim_{\wt_{\infty} \to \infty} \left[ \frac{\wt_{\tau}^{2}}{\wt_{\infty}^{2}} \exp\left( + \int_{\wt_{\tau}}^{\wt_{\infty}} \frac{d\wt}{\wt}  \frac{1+f(\wt)}{a(\wt)}\right) \right]^{\frac{2}{3}}
\end{eqnarray}
such that for asymptotically late times $\wt_{\tau}\to \infty$ the function $\mathcal{E}(\wt_{\tau})$ approaches unity. We note that in extending the calculation of~\cite{Giacalone:2019ldn} for charge neutral plasmas, to finite density systems it is important to preserve the correspondence between the scaling variable $\tilde{\omega}$ and the energy density.

\bibliography{bib/ref}

\begin{thebibliography}{72}
\expandafter\ifx\csname natexlab\endcsname\relax\def\natexlab#1{#1}\fi
\expandafter\ifx\csname bibnamefont\endcsname\relax
  \def\bibnamefont#1{#1}\fi
\expandafter\ifx\csname bibfnamefont\endcsname\relax
  \def\bibfnamefont#1{#1}\fi
\expandafter\ifx\csname citenamefont\endcsname\relax
  \def\citenamefont#1{#1}\fi
\expandafter\ifx\csname url\endcsname\relax
  \def\url#1{\texttt{#1}}\fi
\expandafter\ifx\csname urlprefix\endcsname\relax\def\urlprefix{URL }\fi
\providecommand{\bibinfo}[2]{#2}
\providecommand{\eprint}[2][]{\url{#2}}

\bibitem[{\citenamefont{Schlichting and Teaney}(2019)}]{Schlichting:2019abc}
\bibinfo{author}{\bibfnamefont{S.}~\bibnamefont{Schlichting}} \bibnamefont{and}
  \bibinfo{author}{\bibfnamefont{D.}~\bibnamefont{Teaney}},
  \bibinfo{journal}{Ann. Rev. Nucl. Part. Sci.} \textbf{\bibinfo{volume}{69}},
  \bibinfo{pages}{447} (\bibinfo{year}{2019}), \eprint{1908.02113}.

\bibitem[{\citenamefont{Berges et~al.}(2020)\citenamefont{Berges, Heller,
  Mazeliauskas, and Venugopalan}}]{Berges:2020fwq}
\bibinfo{author}{\bibfnamefont{J.}~\bibnamefont{Berges}},
  \bibinfo{author}{\bibfnamefont{M.~P.} \bibnamefont{Heller}},
  \bibinfo{author}{\bibfnamefont{A.}~\bibnamefont{Mazeliauskas}},
  \bibnamefont{and}
  \bibinfo{author}{\bibfnamefont{R.}~\bibnamefont{Venugopalan}}
  (\bibinfo{year}{2020}), \eprint{2005.12299}.

\bibitem[{\citenamefont{Gale et~al.}(2013)\citenamefont{Gale, Jeon, and
  Schenke}}]{Gale:2013da}
\bibinfo{author}{\bibfnamefont{C.}~\bibnamefont{Gale}},
  \bibinfo{author}{\bibfnamefont{S.}~\bibnamefont{Jeon}}, \bibnamefont{and}
  \bibinfo{author}{\bibfnamefont{B.}~\bibnamefont{Schenke}},
  \bibinfo{journal}{Int. J. Mod. Phys. A} \textbf{\bibinfo{volume}{28}},
  \bibinfo{pages}{1340011} (\bibinfo{year}{2013}), \eprint{1301.5893}.

\bibitem[{\citenamefont{Heinz and Snellings}(2013)}]{Heinz:2013th}
\bibinfo{author}{\bibfnamefont{U.}~\bibnamefont{Heinz}} \bibnamefont{and}
  \bibinfo{author}{\bibfnamefont{R.}~\bibnamefont{Snellings}},
  \bibinfo{journal}{Ann. Rev. Nucl. Part. Sci.} \textbf{\bibinfo{volume}{63}},
  \bibinfo{pages}{123} (\bibinfo{year}{2013}), \eprint{1301.2826}.

\bibitem[{\citenamefont{Chesler and Yaffe}(2009)}]{Chesler:2008hg}
\bibinfo{author}{\bibfnamefont{P.~M.} \bibnamefont{Chesler}} \bibnamefont{and}
  \bibinfo{author}{\bibfnamefont{L.~G.} \bibnamefont{Yaffe}},
  \bibinfo{journal}{Phys. Rev. Lett.} \textbf{\bibinfo{volume}{102}},
  \bibinfo{pages}{211601} (\bibinfo{year}{2009}), \eprint{0812.2053}.

\bibitem[{\citenamefont{Balasubramanian
  et~al.}(2011)\citenamefont{Balasubramanian, Bernamonti, de~Boer, Copland,
  Craps, Keski-Vakkuri, Muller, Schafer, Shigemori, and
  Staessens}}]{Balasubramanian:2010ce}
\bibinfo{author}{\bibfnamefont{V.}~\bibnamefont{Balasubramanian}},
  \bibinfo{author}{\bibfnamefont{A.}~\bibnamefont{Bernamonti}},
  \bibinfo{author}{\bibfnamefont{J.}~\bibnamefont{de~Boer}},
  \bibinfo{author}{\bibfnamefont{N.}~\bibnamefont{Copland}},
  \bibinfo{author}{\bibfnamefont{B.}~\bibnamefont{Craps}},
  \bibinfo{author}{\bibfnamefont{E.}~\bibnamefont{Keski-Vakkuri}},
  \bibinfo{author}{\bibfnamefont{B.}~\bibnamefont{Muller}},
  \bibinfo{author}{\bibfnamefont{A.}~\bibnamefont{Schafer}},
  \bibinfo{author}{\bibfnamefont{M.}~\bibnamefont{Shigemori}},
  \bibnamefont{and}
  \bibinfo{author}{\bibfnamefont{W.}~\bibnamefont{Staessens}},
  \bibinfo{journal}{Phys. Rev. Lett.} \textbf{\bibinfo{volume}{106}},
  \bibinfo{pages}{191601} (\bibinfo{year}{2011}), \eprint{1012.4753}.

\bibitem[{\citenamefont{Heller et~al.}(2012)\citenamefont{Heller, Janik, and
  Witaszczyk}}]{Heller:2011ju}
\bibinfo{author}{\bibfnamefont{M.~P.} \bibnamefont{Heller}},
  \bibinfo{author}{\bibfnamefont{R.~A.} \bibnamefont{Janik}}, \bibnamefont{and}
  \bibinfo{author}{\bibfnamefont{P.}~\bibnamefont{Witaszczyk}},
  \bibinfo{journal}{Phys. Rev. Lett.} \textbf{\bibinfo{volume}{108}},
  \bibinfo{pages}{201602} (\bibinfo{year}{2012}), \eprint{1103.3452}.

\bibitem[{\citenamefont{Keegan et~al.}(2016)\citenamefont{Keegan, Kurkela,
  Romatschke, van~der Schee, and Zhu}}]{Keegan:2015avk}
\bibinfo{author}{\bibfnamefont{L.}~\bibnamefont{Keegan}},
  \bibinfo{author}{\bibfnamefont{A.}~\bibnamefont{Kurkela}},
  \bibinfo{author}{\bibfnamefont{P.}~\bibnamefont{Romatschke}},
  \bibinfo{author}{\bibfnamefont{W.}~\bibnamefont{van~der Schee}},
  \bibnamefont{and} \bibinfo{author}{\bibfnamefont{Y.}~\bibnamefont{Zhu}},
  \bibinfo{journal}{JHEP} \textbf{\bibinfo{volume}{04}}, \bibinfo{pages}{031}
  (\bibinfo{year}{2016}), \eprint{1512.05347}.

\bibitem[{\citenamefont{Kurkela and Lu}(2014)}]{Kurkela:2014tea}
\bibinfo{author}{\bibfnamefont{A.}~\bibnamefont{Kurkela}} \bibnamefont{and}
  \bibinfo{author}{\bibfnamefont{E.}~\bibnamefont{Lu}}, \bibinfo{journal}{Phys.
  Rev. Lett.} \textbf{\bibinfo{volume}{113}}, \bibinfo{pages}{182301}
  (\bibinfo{year}{2014}), \eprint{1405.6318}.

\bibitem[{\citenamefont{Kurkela
  et~al.}(2019{\natexlab{a}})\citenamefont{Kurkela, Mazeliauskas, Paquet,
  Schlichting, and Teaney}}]{Kurkela:2018wud}
\bibinfo{author}{\bibfnamefont{A.}~\bibnamefont{Kurkela}},
  \bibinfo{author}{\bibfnamefont{A.}~\bibnamefont{Mazeliauskas}},
  \bibinfo{author}{\bibfnamefont{J.-F.} \bibnamefont{Paquet}},
  \bibinfo{author}{\bibfnamefont{S.}~\bibnamefont{Schlichting}},
  \bibnamefont{and} \bibinfo{author}{\bibfnamefont{D.}~\bibnamefont{Teaney}},
  \bibinfo{journal}{Phys. Rev. Lett.} \textbf{\bibinfo{volume}{122}},
  \bibinfo{pages}{122302} (\bibinfo{year}{2019}{\natexlab{a}}),
  \eprint{1805.01604}.

\bibitem[{\citenamefont{Heller et~al.}(2018)\citenamefont{Heller, Kurkela,
  Spali\'nski, and Svensson}}]{Heller:2016rtz}
\bibinfo{author}{\bibfnamefont{M.~P.} \bibnamefont{Heller}},
  \bibinfo{author}{\bibfnamefont{A.}~\bibnamefont{Kurkela}},
  \bibinfo{author}{\bibfnamefont{M.}~\bibnamefont{Spali\'nski}},
  \bibnamefont{and} \bibinfo{author}{\bibfnamefont{V.}~\bibnamefont{Svensson}},
  \bibinfo{journal}{Phys. Rev. D} \textbf{\bibinfo{volume}{97}},
  \bibinfo{pages}{091503} (\bibinfo{year}{2018}), \eprint{1609.04803}.

\bibitem[{\citenamefont{Almaalol et~al.}(2020)\citenamefont{Almaalol, Kurkela,
  and Strickland}}]{Almaalol:2020rnu}
\bibinfo{author}{\bibfnamefont{D.}~\bibnamefont{Almaalol}},
  \bibinfo{author}{\bibfnamefont{A.}~\bibnamefont{Kurkela}}, \bibnamefont{and}
  \bibinfo{author}{\bibfnamefont{M.}~\bibnamefont{Strickland}},
  \bibinfo{journal}{Phys. Rev. Lett.} \textbf{\bibinfo{volume}{125}},
  \bibinfo{pages}{122302} (\bibinfo{year}{2020}), \eprint{2004.05195}.

\bibitem[{\citenamefont{Kurkela and Mazeliauskas}(2019)}]{Kurkela:2018oqw}
\bibinfo{author}{\bibfnamefont{A.}~\bibnamefont{Kurkela}} \bibnamefont{and}
  \bibinfo{author}{\bibfnamefont{A.}~\bibnamefont{Mazeliauskas}},
  \bibinfo{journal}{Phys. Rev. D} \textbf{\bibinfo{volume}{99}},
  \bibinfo{pages}{054018} (\bibinfo{year}{2019}), \eprint{1811.03068}.

\bibitem[{\citenamefont{Blaizot and Yan}(2020{\natexlab{a}})}]{Blaizot:2020gql}
\bibinfo{author}{\bibfnamefont{J.-P.} \bibnamefont{Blaizot}} \bibnamefont{and}
  \bibinfo{author}{\bibfnamefont{L.}~\bibnamefont{Yan}}
  (\bibinfo{year}{2020}{\natexlab{a}}), \eprint{2006.08815}.

\bibitem[{\citenamefont{Blaizot and Yan}(2020{\natexlab{b}})}]{Blaizot:2019scw}
\bibinfo{author}{\bibfnamefont{J.-P.} \bibnamefont{Blaizot}} \bibnamefont{and}
  \bibinfo{author}{\bibfnamefont{L.}~\bibnamefont{Yan}},
  \bibinfo{journal}{Annals Phys.} \textbf{\bibinfo{volume}{412}},
  \bibinfo{pages}{167993} (\bibinfo{year}{2020}{\natexlab{b}}),
  \eprint{1904.08677}.

\bibitem[{\citenamefont{Blaizot and Yan}(2018)}]{Blaizot:2017ucy}
\bibinfo{author}{\bibfnamefont{J.-P.} \bibnamefont{Blaizot}} \bibnamefont{and}
  \bibinfo{author}{\bibfnamefont{L.}~\bibnamefont{Yan}},
  \bibinfo{journal}{Phys. Lett. B} \textbf{\bibinfo{volume}{780}},
  \bibinfo{pages}{283} (\bibinfo{year}{2018}), \eprint{1712.03856}.

\bibitem[{\citenamefont{Kamata et~al.}(2020)\citenamefont{Kamata, Martinez,
  Plaschke, Ochsenfeld, and Schlichting}}]{Kamata:2020mka}
\bibinfo{author}{\bibfnamefont{S.}~\bibnamefont{Kamata}},
  \bibinfo{author}{\bibfnamefont{M.}~\bibnamefont{Martinez}},
  \bibinfo{author}{\bibfnamefont{P.}~\bibnamefont{Plaschke}},
  \bibinfo{author}{\bibfnamefont{S.}~\bibnamefont{Ochsenfeld}},
  \bibnamefont{and}
  \bibinfo{author}{\bibfnamefont{S.}~\bibnamefont{Schlichting}},
  \bibinfo{journal}{Phys. Rev. D} \textbf{\bibinfo{volume}{102}},
  \bibinfo{pages}{056003} (\bibinfo{year}{2020}), \eprint{2004.06751}.

\bibitem[{\citenamefont{Strickland}(2018)}]{Strickland:2018ayk}
\bibinfo{author}{\bibfnamefont{M.}~\bibnamefont{Strickland}},
  \bibinfo{journal}{JHEP} \textbf{\bibinfo{volume}{12}}, \bibinfo{pages}{128}
  (\bibinfo{year}{2018}), \eprint{1809.01200}.

\bibitem[{\citenamefont{Martinez et~al.}(2012)\citenamefont{Martinez,
  Ryblewski, and Strickland}}]{Martinez:2012tu}
\bibinfo{author}{\bibfnamefont{M.}~\bibnamefont{Martinez}},
  \bibinfo{author}{\bibfnamefont{R.}~\bibnamefont{Ryblewski}},
  \bibnamefont{and}
  \bibinfo{author}{\bibfnamefont{M.}~\bibnamefont{Strickland}},
  \bibinfo{journal}{Phys. Rev. C} \textbf{\bibinfo{volume}{85}},
  \bibinfo{pages}{064913} (\bibinfo{year}{2012}), \eprint{1204.1473}.

\bibitem[{\citenamefont{Kurkela et~al.}(2020)\citenamefont{Kurkela, van~der
  Schee, Wiedemann, and Wu}}]{Kurkela:2019set}
\bibinfo{author}{\bibfnamefont{A.}~\bibnamefont{Kurkela}},
  \bibinfo{author}{\bibfnamefont{W.}~\bibnamefont{van~der Schee}},
  \bibinfo{author}{\bibfnamefont{U.~A.} \bibnamefont{Wiedemann}},
  \bibnamefont{and} \bibinfo{author}{\bibfnamefont{B.}~\bibnamefont{Wu}},
  \bibinfo{journal}{Phys. Rev. Lett.} \textbf{\bibinfo{volume}{124}},
  \bibinfo{pages}{102301} (\bibinfo{year}{2020}), \eprint{1907.08101}.

\bibitem[{\citenamefont{Behtash et~al.}(2020)\citenamefont{Behtash, Kamata,
  Martinez, Sch\"afer, and Skokov}}]{Behtash:2020vqk}
\bibinfo{author}{\bibfnamefont{A.}~\bibnamefont{Behtash}},
  \bibinfo{author}{\bibfnamefont{S.}~\bibnamefont{Kamata}},
  \bibinfo{author}{\bibfnamefont{M.}~\bibnamefont{Martinez}},
  \bibinfo{author}{\bibfnamefont{T.}~\bibnamefont{Sch\"afer}},
  \bibnamefont{and} \bibinfo{author}{\bibfnamefont{V.}~\bibnamefont{Skokov}}
  (\bibinfo{year}{2020}), \eprint{2011.08235}.

\bibitem[{\citenamefont{Heller and Spalinski}(2015)}]{Heller:2015dha}
\bibinfo{author}{\bibfnamefont{M.~P.} \bibnamefont{Heller}} \bibnamefont{and}
  \bibinfo{author}{\bibfnamefont{M.}~\bibnamefont{Spalinski}},
  \bibinfo{journal}{Phys. Rev. Lett.} \textbf{\bibinfo{volume}{115}},
  \bibinfo{pages}{072501} (\bibinfo{year}{2015}), \eprint{1503.07514}.

\bibitem[{\citenamefont{Romatschke}(2018)}]{Romatschke:2017vte}
\bibinfo{author}{\bibfnamefont{P.}~\bibnamefont{Romatschke}},
  \bibinfo{journal}{Phys. Rev. Lett.} \textbf{\bibinfo{volume}{120}},
  \bibinfo{pages}{012301} (\bibinfo{year}{2018}), \eprint{1704.08699}.

\bibitem[{\citenamefont{Behtash et~al.}(2019)\citenamefont{Behtash, Kamata,
  Martinez, and Shi}}]{Behtash:2019txb}
\bibinfo{author}{\bibfnamefont{A.}~\bibnamefont{Behtash}},
  \bibinfo{author}{\bibfnamefont{S.}~\bibnamefont{Kamata}},
  \bibinfo{author}{\bibfnamefont{M.}~\bibnamefont{Martinez}}, \bibnamefont{and}
  \bibinfo{author}{\bibfnamefont{H.}~\bibnamefont{Shi}},
  \bibinfo{journal}{Phys. Rev. D} \textbf{\bibinfo{volume}{99}},
  \bibinfo{pages}{116012} (\bibinfo{year}{2019}), \eprint{1901.08632}.

\bibitem[{\citenamefont{Heller et~al.}(2020)\citenamefont{Heller, Jefferson,
  Spali\'nski, and Svensson}}]{Heller:2020anv}
\bibinfo{author}{\bibfnamefont{M.~P.} \bibnamefont{Heller}},
  \bibinfo{author}{\bibfnamefont{R.}~\bibnamefont{Jefferson}},
  \bibinfo{author}{\bibfnamefont{M.}~\bibnamefont{Spali\'nski}},
  \bibnamefont{and} \bibinfo{author}{\bibfnamefont{V.}~\bibnamefont{Svensson}},
  \bibinfo{journal}{Phys. Rev. Lett.} \textbf{\bibinfo{volume}{125}},
  \bibinfo{pages}{132301} (\bibinfo{year}{2020}), \eprint{2003.07368}.

\bibitem[{\citenamefont{Denicol and Noronha}(2020)}]{Denicol:2019lio}
\bibinfo{author}{\bibfnamefont{G.~S.} \bibnamefont{Denicol}} \bibnamefont{and}
  \bibinfo{author}{\bibfnamefont{J.}~\bibnamefont{Noronha}},
  \bibinfo{journal}{Phys. Rev. Lett.} \textbf{\bibinfo{volume}{124}},
  \bibinfo{pages}{152301} (\bibinfo{year}{2020}), \eprint{1908.09957}.

\bibitem[{\citenamefont{Strickland et~al.}(2018)\citenamefont{Strickland,
  Noronha, and Denicol}}]{Strickland:2017kux}
\bibinfo{author}{\bibfnamefont{M.}~\bibnamefont{Strickland}},
  \bibinfo{author}{\bibfnamefont{J.}~\bibnamefont{Noronha}}, \bibnamefont{and}
  \bibinfo{author}{\bibfnamefont{G.}~\bibnamefont{Denicol}},
  \bibinfo{journal}{Phys. Rev. D} \textbf{\bibinfo{volume}{97}},
  \bibinfo{pages}{036020} (\bibinfo{year}{2018}), \eprint{1709.06644}.

\bibitem[{\citenamefont{Giacalone et~al.}(2019)\citenamefont{Giacalone,
  Mazeliauskas, and Schlichting}}]{Giacalone:2019ldn}
\bibinfo{author}{\bibfnamefont{G.}~\bibnamefont{Giacalone}},
  \bibinfo{author}{\bibfnamefont{A.}~\bibnamefont{Mazeliauskas}},
  \bibnamefont{and}
  \bibinfo{author}{\bibfnamefont{S.}~\bibnamefont{Schlichting}},
  \bibinfo{journal}{Phys. Rev. Lett.} \textbf{\bibinfo{volume}{123}},
  \bibinfo{pages}{262301} (\bibinfo{year}{2019}), \eprint{1908.02866}.

\bibitem[{\citenamefont{Romatschke}(2017)}]{Romatschke:2017acs}
\bibinfo{author}{\bibfnamefont{P.}~\bibnamefont{Romatschke}},
  \bibinfo{journal}{JHEP} \textbf{\bibinfo{volume}{12}}, \bibinfo{pages}{079}
  (\bibinfo{year}{2017}), \eprint{1710.03234}.

\bibitem[{\citenamefont{Denicol and Noronha}(2018)}]{Denicol:2017lxn}
\bibinfo{author}{\bibfnamefont{G.~S.} \bibnamefont{Denicol}} \bibnamefont{and}
  \bibinfo{author}{\bibfnamefont{J.}~\bibnamefont{Noronha}},
  \bibinfo{journal}{Phys. Rev. D} \textbf{\bibinfo{volume}{97}},
  \bibinfo{pages}{056021} (\bibinfo{year}{2018}), \eprint{1711.01657}.

\bibitem[{\citenamefont{Kurkela
  et~al.}(2019{\natexlab{b}})\citenamefont{Kurkela, Mazeliauskas, Paquet,
  Schlichting, and Teaney}}]{Kurkela:2018vqr}
\bibinfo{author}{\bibfnamefont{A.}~\bibnamefont{Kurkela}},
  \bibinfo{author}{\bibfnamefont{A.}~\bibnamefont{Mazeliauskas}},
  \bibinfo{author}{\bibfnamefont{J.-F.} \bibnamefont{Paquet}},
  \bibinfo{author}{\bibfnamefont{S.}~\bibnamefont{Schlichting}},
  \bibnamefont{and} \bibinfo{author}{\bibfnamefont{D.}~\bibnamefont{Teaney}},
  \bibinfo{journal}{Phys. Rev. C} \textbf{\bibinfo{volume}{99}},
  \bibinfo{pages}{034910} (\bibinfo{year}{2019}{\natexlab{b}}),
  \eprint{1805.00961}.

\bibitem[{\citenamefont{Nunes~da Silva et~al.}(2020)\citenamefont{Nunes~da
  Silva, Chinellato, Hippert, Serenone, Takahashi, Denicol, Luzum, and
  Noronha}}]{NunesdaSilva:2020bfs}
\bibinfo{author}{\bibfnamefont{T.}~\bibnamefont{Nunes~da Silva}},
  \bibinfo{author}{\bibfnamefont{D.}~\bibnamefont{Chinellato}},
  \bibinfo{author}{\bibfnamefont{M.}~\bibnamefont{Hippert}},
  \bibinfo{author}{\bibfnamefont{W.}~\bibnamefont{Serenone}},
  \bibinfo{author}{\bibfnamefont{J.}~\bibnamefont{Takahashi}},
  \bibinfo{author}{\bibfnamefont{G.~S.} \bibnamefont{Denicol}},
  \bibinfo{author}{\bibfnamefont{M.}~\bibnamefont{Luzum}}, \bibnamefont{and}
  \bibinfo{author}{\bibfnamefont{J.}~\bibnamefont{Noronha}}
  (\bibinfo{year}{2020}), \eprint{2006.02324}.

\bibitem[{\citenamefont{Gale et~al.}(2020)\citenamefont{Gale, Paquet, Schenke,
  and Shen}}]{Gale:2020xlg}
\bibinfo{author}{\bibfnamefont{C.}~\bibnamefont{Gale}},
  \bibinfo{author}{\bibfnamefont{J.-F.} \bibnamefont{Paquet}},
  \bibinfo{author}{\bibfnamefont{B.}~\bibnamefont{Schenke}}, \bibnamefont{and}
  \bibinfo{author}{\bibfnamefont{C.}~\bibnamefont{Shen}}, in
  \emph{\bibinfo{booktitle}{{28th International Conference on Ultrarelativistic
  Nucleus-Nucleus Collisions}}} (\bibinfo{year}{2020}), \eprint{2002.05191}.

\bibitem[{\citenamefont{Arnold et~al.}(2003)\citenamefont{Arnold, Moore, and
  Yaffe}}]{Arnold:2002zm}
\bibinfo{author}{\bibfnamefont{P.~B.} \bibnamefont{Arnold}},
  \bibinfo{author}{\bibfnamefont{G.~D.} \bibnamefont{Moore}}, \bibnamefont{and}
  \bibinfo{author}{\bibfnamefont{L.~G.} \bibnamefont{Yaffe}},
  \bibinfo{journal}{JHEP} \textbf{\bibinfo{volume}{01}}, \bibinfo{pages}{030}
  (\bibinfo{year}{2003}), \eprint{hep-ph/0209353}.

\bibitem[{\citenamefont{Du and Schlichting}(2020)}]{Du:2020pre}
\bibinfo{author}{\bibfnamefont{X.}~\bibnamefont{Du}} \bibnamefont{and}
  \bibinfo{author}{\bibfnamefont{S.}~\bibnamefont{Schlichting}},
  \bibinfo{journal}{in preparation}  (\bibinfo{year}{2020}).

\bibitem[{\citenamefont{Gelis et~al.}(2010)\citenamefont{Gelis, Iancu,
  Jalilian-Marian, and Venugopalan}}]{Gelis:2010nm}
\bibinfo{author}{\bibfnamefont{F.}~\bibnamefont{Gelis}},
  \bibinfo{author}{\bibfnamefont{E.}~\bibnamefont{Iancu}},
  \bibinfo{author}{\bibfnamefont{J.}~\bibnamefont{Jalilian-Marian}},
  \bibnamefont{and}
  \bibinfo{author}{\bibfnamefont{R.}~\bibnamefont{Venugopalan}},
  \bibinfo{journal}{Ann. Rev. Nucl. Part. Sci.} \textbf{\bibinfo{volume}{60}},
  \bibinfo{pages}{463} (\bibinfo{year}{2010}), \eprint{1002.0333}.

\bibitem[{\citenamefont{Iancu and Venugopalan}(2003)}]{Iancu:2003xm}
\bibinfo{author}{\bibfnamefont{E.}~\bibnamefont{Iancu}} \bibnamefont{and}
  \bibinfo{author}{\bibfnamefont{R.}~\bibnamefont{Venugopalan}},
  \emph{\bibinfo{title}{{The Color glass condensate and high-energy scattering
  in QCD}}} (\bibinfo{year}{2003}), pp. \bibinfo{pages}{249--3363},
  \eprint{hep-ph/0303204}.

\bibitem[{\citenamefont{Kovner et~al.}(1995)\citenamefont{Kovner, McLerran, and
  Weigert}}]{Kovner:1995ja}
\bibinfo{author}{\bibfnamefont{A.}~\bibnamefont{Kovner}},
  \bibinfo{author}{\bibfnamefont{L.~D.} \bibnamefont{McLerran}},
  \bibnamefont{and} \bibinfo{author}{\bibfnamefont{H.}~\bibnamefont{Weigert}},
  \bibinfo{journal}{Phys. Rev. D} \textbf{\bibinfo{volume}{52}},
  \bibinfo{pages}{6231} (\bibinfo{year}{1995}), \eprint{hep-ph/9502289}.

\bibitem[{\citenamefont{Krasnitz and Venugopalan}(1999)}]{Krasnitz:1998ns}
\bibinfo{author}{\bibfnamefont{A.}~\bibnamefont{Krasnitz}} \bibnamefont{and}
  \bibinfo{author}{\bibfnamefont{R.}~\bibnamefont{Venugopalan}},
  \bibinfo{journal}{Nucl. Phys. B} \textbf{\bibinfo{volume}{557}},
  \bibinfo{pages}{237} (\bibinfo{year}{1999}), \eprint{hep-ph/9809433}.

\bibitem[{\citenamefont{Krasnitz et~al.}(2003)\citenamefont{Krasnitz, Nara, and
  Venugopalan}}]{Krasnitz:2003jw}
\bibinfo{author}{\bibfnamefont{A.}~\bibnamefont{Krasnitz}},
  \bibinfo{author}{\bibfnamefont{Y.}~\bibnamefont{Nara}}, \bibnamefont{and}
  \bibinfo{author}{\bibfnamefont{R.}~\bibnamefont{Venugopalan}},
  \bibinfo{journal}{Nucl. Phys. A} \textbf{\bibinfo{volume}{727}},
  \bibinfo{pages}{427} (\bibinfo{year}{2003}), \eprint{hep-ph/0305112}.

\bibitem[{\citenamefont{Blaizot et~al.}(2010)\citenamefont{Blaizot, Lappi, and
  Mehtar-Tani}}]{Blaizot:2010kh}
\bibinfo{author}{\bibfnamefont{J.-P.} \bibnamefont{Blaizot}},
  \bibinfo{author}{\bibfnamefont{T.}~\bibnamefont{Lappi}}, \bibnamefont{and}
  \bibinfo{author}{\bibfnamefont{Y.}~\bibnamefont{Mehtar-Tani}},
  \bibinfo{journal}{Nucl. Phys. A} \textbf{\bibinfo{volume}{846}},
  \bibinfo{pages}{63} (\bibinfo{year}{2010}), \eprint{1005.0955}.

\bibitem[{\citenamefont{Schenke et~al.}(2012)\citenamefont{Schenke, Tribedy,
  and Venugopalan}}]{Schenke:2012hg}
\bibinfo{author}{\bibfnamefont{B.}~\bibnamefont{Schenke}},
  \bibinfo{author}{\bibfnamefont{P.}~\bibnamefont{Tribedy}}, \bibnamefont{and}
  \bibinfo{author}{\bibfnamefont{R.}~\bibnamefont{Venugopalan}},
  \bibinfo{journal}{Phys. Rev. C} \textbf{\bibinfo{volume}{86}},
  \bibinfo{pages}{034908} (\bibinfo{year}{2012}), \eprint{1206.6805}.

\bibitem[{\citenamefont{Gelis and Tanji}(2016)}]{Gelis:2015eua}
\bibinfo{author}{\bibfnamefont{F.}~\bibnamefont{Gelis}} \bibnamefont{and}
  \bibinfo{author}{\bibfnamefont{N.}~\bibnamefont{Tanji}},
  \bibinfo{journal}{JHEP} \textbf{\bibinfo{volume}{02}}, \bibinfo{pages}{126}
  (\bibinfo{year}{2016}), \eprint{1506.03327}.

\bibitem[{\citenamefont{Fujii et~al.}(2006)\citenamefont{Fujii, Gelis, and
  Venugopalan}}]{Fujii:2006ab}
\bibinfo{author}{\bibfnamefont{H.}~\bibnamefont{Fujii}},
  \bibinfo{author}{\bibfnamefont{F.}~\bibnamefont{Gelis}}, \bibnamefont{and}
  \bibinfo{author}{\bibfnamefont{R.}~\bibnamefont{Venugopalan}},
  \bibinfo{journal}{Nucl. Phys. A} \textbf{\bibinfo{volume}{780}},
  \bibinfo{pages}{146} (\bibinfo{year}{2006}), \eprint{hep-ph/0603099}.

\bibitem[{\citenamefont{Gelis et~al.}(2006)\citenamefont{Gelis, Kajantie, and
  Lappi}}]{Gelis:2005pb}
\bibinfo{author}{\bibfnamefont{F.}~\bibnamefont{Gelis}},
  \bibinfo{author}{\bibfnamefont{K.}~\bibnamefont{Kajantie}}, \bibnamefont{and}
  \bibinfo{author}{\bibfnamefont{T.}~\bibnamefont{Lappi}},
  \bibinfo{journal}{Phys. Rev. Lett.} \textbf{\bibinfo{volume}{96}},
  \bibinfo{pages}{032304} (\bibinfo{year}{2006}), \eprint{hep-ph/0508229}.

\bibitem[{\citenamefont{McLerran et~al.}(2019)\citenamefont{McLerran,
  Schlichting, and Sen}}]{McLerran:2018avb}
\bibinfo{author}{\bibfnamefont{L.~D.} \bibnamefont{McLerran}},
  \bibinfo{author}{\bibfnamefont{S.}~\bibnamefont{Schlichting}},
  \bibnamefont{and} \bibinfo{author}{\bibfnamefont{S.}~\bibnamefont{Sen}},
  \bibinfo{journal}{Phys. Rev. D} \textbf{\bibinfo{volume}{99}},
  \bibinfo{pages}{074009} (\bibinfo{year}{2019}), \eprint{1811.04089}.

\bibitem[{\citenamefont{Lushozi et~al.}(2020)\citenamefont{Lushozi, McLerran,
  Praszalowicz, and Yu}}]{Lushozi:2019duv}
\bibinfo{author}{\bibfnamefont{M.}~\bibnamefont{Lushozi}},
  \bibinfo{author}{\bibfnamefont{L.~D.} \bibnamefont{McLerran}},
  \bibinfo{author}{\bibfnamefont{M.}~\bibnamefont{Praszalowicz}},
  \bibnamefont{and} \bibinfo{author}{\bibfnamefont{G.}~\bibnamefont{Yu}},
  \bibinfo{journal}{Phys. Rev. C} \textbf{\bibinfo{volume}{102}},
  \bibinfo{pages}{034908} (\bibinfo{year}{2020}), \eprint{1912.08553}.

\bibitem[{\citenamefont{Kajantie et~al.}(2020)\citenamefont{Kajantie, McLerran,
  and Paatelainen}}]{Kajantie:2019nse}
\bibinfo{author}{\bibfnamefont{K.}~\bibnamefont{Kajantie}},
  \bibinfo{author}{\bibfnamefont{L.~D.} \bibnamefont{McLerran}},
  \bibnamefont{and}
  \bibinfo{author}{\bibfnamefont{R.}~\bibnamefont{Paatelainen}},
  \bibinfo{journal}{Phys. Rev. D} \textbf{\bibinfo{volume}{101}},
  \bibinfo{pages}{054012} (\bibinfo{year}{2020}), \eprint{1911.12738}.

\bibitem[{\citenamefont{Agostini et~al.}(2019)\citenamefont{Agostini,
  Altinoluk, and Armesto}}]{Agostini:2019avp}
\bibinfo{author}{\bibfnamefont{P.}~\bibnamefont{Agostini}},
  \bibinfo{author}{\bibfnamefont{T.}~\bibnamefont{Altinoluk}},
  \bibnamefont{and} \bibinfo{author}{\bibfnamefont{N.}~\bibnamefont{Armesto}},
  \bibinfo{journal}{Eur. Phys. J. C} \textbf{\bibinfo{volume}{79}},
  \bibinfo{pages}{600} (\bibinfo{year}{2019}), \eprint{1902.04483}.

\bibitem[{\citenamefont{Mueller}(2000)}]{Mueller:1999pi}
\bibinfo{author}{\bibfnamefont{A.~H.} \bibnamefont{Mueller}},
  \bibinfo{journal}{Phys. Lett. B} \textbf{\bibinfo{volume}{475}},
  \bibinfo{pages}{220} (\bibinfo{year}{2000}), \eprint{hep-ph/9909388}.

\bibitem[{\citenamefont{Landau and
  Pomeranchuk}(1953{\natexlab{a}})}]{Landau:1953gr}
\bibinfo{author}{\bibfnamefont{L.}~\bibnamefont{Landau}} \bibnamefont{and}
  \bibinfo{author}{\bibfnamefont{I.}~\bibnamefont{Pomeranchuk}},
  \bibinfo{journal}{Dokl. Akad. Nauk Ser. Fiz.} \textbf{\bibinfo{volume}{92}},
  \bibinfo{pages}{735} (\bibinfo{year}{1953}{\natexlab{a}}).

\bibitem[{\citenamefont{Landau and
  Pomeranchuk}(1953{\natexlab{b}})}]{Landau:1953um}
\bibinfo{author}{\bibfnamefont{L.}~\bibnamefont{Landau}} \bibnamefont{and}
  \bibinfo{author}{\bibfnamefont{I.}~\bibnamefont{Pomeranchuk}},
  \bibinfo{journal}{Dokl. Akad. Nauk Ser. Fiz.} \textbf{\bibinfo{volume}{92}},
  \bibinfo{pages}{535} (\bibinfo{year}{1953}{\natexlab{b}}).

\bibitem[{\citenamefont{Migdal}(1955)}]{Migdal:1955nv}
\bibinfo{author}{\bibfnamefont{A.~B.} \bibnamefont{Migdal}},
  \bibinfo{journal}{Dokl. Akad. Nauk Ser. Fiz.} \textbf{\bibinfo{volume}{105}},
  \bibinfo{pages}{77} (\bibinfo{year}{1955}).

\bibitem[{\citenamefont{Biro and Zimanyi}(1982)}]{Biro:1981zi}
\bibinfo{author}{\bibfnamefont{T.}~\bibnamefont{Biro}} \bibnamefont{and}
  \bibinfo{author}{\bibfnamefont{J.}~\bibnamefont{Zimanyi}},
  \bibinfo{journal}{Phys. Lett. B} \textbf{\bibinfo{volume}{113}},
  \bibinfo{pages}{6} (\bibinfo{year}{1982}).

\bibitem[{\citenamefont{Bjorken}(1983)}]{Bjorken:1982qr}
\bibinfo{author}{\bibfnamefont{J.}~\bibnamefont{Bjorken}},
  \bibinfo{journal}{Phys. Rev. D} \textbf{\bibinfo{volume}{27}},
  \bibinfo{pages}{140} (\bibinfo{year}{1983}).

\bibitem[{\citenamefont{Gyulassy and Matsui}(1984)}]{Gyulassy:1983ub}
\bibinfo{author}{\bibfnamefont{M.}~\bibnamefont{Gyulassy}} \bibnamefont{and}
  \bibinfo{author}{\bibfnamefont{T.}~\bibnamefont{Matsui}},
  \bibinfo{journal}{Phys. Rev. D} \textbf{\bibinfo{volume}{29}},
  \bibinfo{pages}{419} (\bibinfo{year}{1984}).

\bibitem[{\citenamefont{Jankowski et~al.}(2020)\citenamefont{Jankowski, Kamata,
  Martinez, and Spali\'nski}}]{Jankowski:2020itt}
\bibinfo{author}{\bibfnamefont{J.}~\bibnamefont{Jankowski}},
  \bibinfo{author}{\bibfnamefont{S.}~\bibnamefont{Kamata}},
  \bibinfo{author}{\bibfnamefont{M.}~\bibnamefont{Martinez}}, \bibnamefont{and}
  \bibinfo{author}{\bibfnamefont{M.}~\bibnamefont{Spali\'nski}}
  (\bibinfo{year}{2020}), \eprint{2012.02184}.

\bibitem[{\citenamefont{Hanus et~al.}(2019)\citenamefont{Hanus, Mazeliauskas,
  and Reygers}}]{Hanus:2019fnc}
\bibinfo{author}{\bibfnamefont{P.}~\bibnamefont{Hanus}},
  \bibinfo{author}{\bibfnamefont{A.}~\bibnamefont{Mazeliauskas}},
  \bibnamefont{and} \bibinfo{author}{\bibfnamefont{K.}~\bibnamefont{Reygers}},
  \bibinfo{journal}{Phys. Rev. C} \textbf{\bibinfo{volume}{100}},
  \bibinfo{pages}{064903} (\bibinfo{year}{2019}), \eprint{1908.02792}.

\bibitem[{\citenamefont{Arsene et~al.}(2005)}]{Arsene:2004fa}
\bibinfo{author}{\bibfnamefont{I.}~\bibnamefont{Arsene}} \bibnamefont{et~al.}
  (\bibinfo{collaboration}{BRAHMS}), \bibinfo{journal}{Nucl. Phys. A}
  \textbf{\bibinfo{volume}{757}}, \bibinfo{pages}{1} (\bibinfo{year}{2005}),
  \eprint{nucl-ex/0410020}.

\bibitem[{\citenamefont{Back et~al.}(2006)}]{Back:2005hs}
\bibinfo{author}{\bibfnamefont{B.}~\bibnamefont{Back}} \bibnamefont{et~al.}
  (\bibinfo{collaboration}{PHOBOS}), \bibinfo{journal}{Phys. Rev. C}
  \textbf{\bibinfo{volume}{74}}, \bibinfo{pages}{021901}
  (\bibinfo{year}{2006}), \eprint{nucl-ex/0509034}.

\bibitem[{\citenamefont{Aamodt et~al.}(2011)}]{Aamodt:2010cz}
\bibinfo{author}{\bibfnamefont{K.}~\bibnamefont{Aamodt}} \bibnamefont{et~al.}
  (\bibinfo{collaboration}{ALICE}), \bibinfo{journal}{Phys. Rev. Lett.}
  \textbf{\bibinfo{volume}{106}}, \bibinfo{pages}{032301}
  (\bibinfo{year}{2011}), \eprint{1012.1657}.

\bibitem[{\citenamefont{Bazavov et~al.}(2019)}]{Bazavov:2018mes}
\bibinfo{author}{\bibfnamefont{A.}~\bibnamefont{Bazavov}} \bibnamefont{et~al.}
  (\bibinfo{collaboration}{HotQCD}), \bibinfo{journal}{Phys. Lett. B}
  \textbf{\bibinfo{volume}{795}}, \bibinfo{pages}{15} (\bibinfo{year}{2019}),
  \eprint{1812.08235}.

\bibitem[{\citenamefont{Romatschke}(2010)}]{Romatschke:2009im}
\bibinfo{author}{\bibfnamefont{P.}~\bibnamefont{Romatschke}},
  \bibinfo{journal}{Int. J. Mod. Phys. E} \textbf{\bibinfo{volume}{19}},
  \bibinfo{pages}{1} (\bibinfo{year}{2010}), \eprint{0902.3663}.

\bibitem[{\citenamefont{Yan}(2018)}]{Yan:2017ivm}
\bibinfo{author}{\bibfnamefont{L.}~\bibnamefont{Yan}}, \bibinfo{journal}{Chin.
  Phys. C} \textbf{\bibinfo{volume}{42}}, \bibinfo{pages}{042001}
  (\bibinfo{year}{2018}), \eprint{1712.04580}.

\bibitem[{\citenamefont{Shen and Yan}(2020)}]{Shen:2020mgh}
\bibinfo{author}{\bibfnamefont{C.}~\bibnamefont{Shen}} \bibnamefont{and}
  \bibinfo{author}{\bibfnamefont{L.}~\bibnamefont{Yan}},
  \bibinfo{journal}{Nucl. Sci. Tech.} \textbf{\bibinfo{volume}{31}},
  \bibinfo{pages}{122} (\bibinfo{year}{2020}), \eprint{2010.12377}.

\bibitem[{\citenamefont{Dore et~al.}(2020)\citenamefont{Dore, Noronha-Hostler,
  and McLaughlin}}]{Dore:2020jye}
\bibinfo{author}{\bibfnamefont{T.}~\bibnamefont{Dore}},
  \bibinfo{author}{\bibfnamefont{J.}~\bibnamefont{Noronha-Hostler}},
  \bibnamefont{and}
  \bibinfo{author}{\bibfnamefont{E.}~\bibnamefont{McLaughlin}},
  \bibinfo{journal}{Phys. Rev. D} \textbf{\bibinfo{volume}{102}},
  \bibinfo{pages}{074017} (\bibinfo{year}{2020}), \eprint{2007.15083}.

\bibitem[{\citenamefont{Berges et~al.}(2014{\natexlab{a}})\citenamefont{Berges,
  Boguslavski, Schlichting, and Venugopalan}}]{Berges:2013eia}
\bibinfo{author}{\bibfnamefont{J.}~\bibnamefont{Berges}},
  \bibinfo{author}{\bibfnamefont{K.}~\bibnamefont{Boguslavski}},
  \bibinfo{author}{\bibfnamefont{S.}~\bibnamefont{Schlichting}},
  \bibnamefont{and}
  \bibinfo{author}{\bibfnamefont{R.}~\bibnamefont{Venugopalan}},
  \bibinfo{journal}{Phys. Rev. D} \textbf{\bibinfo{volume}{89}},
  \bibinfo{pages}{074011} (\bibinfo{year}{2014}{\natexlab{a}}),
  \eprint{1303.5650}.

\bibitem[{\citenamefont{Berges et~al.}(2014{\natexlab{b}})\citenamefont{Berges,
  Boguslavski, Schlichting, and Venugopalan}}]{Berges:2013fga}
\bibinfo{author}{\bibfnamefont{J.}~\bibnamefont{Berges}},
  \bibinfo{author}{\bibfnamefont{K.}~\bibnamefont{Boguslavski}},
  \bibinfo{author}{\bibfnamefont{S.}~\bibnamefont{Schlichting}},
  \bibnamefont{and}
  \bibinfo{author}{\bibfnamefont{R.}~\bibnamefont{Venugopalan}},
  \bibinfo{journal}{Phys. Rev. D} \textbf{\bibinfo{volume}{89}},
  \bibinfo{pages}{114007} (\bibinfo{year}{2014}{\natexlab{b}}),
  \eprint{1311.3005}.

\bibitem[{\citenamefont{Martinez et~al.}(2019)\citenamefont{Martinez, Sievert,
  Wertepny, and Noronha-Hostler}}]{Martinez:2019jbu}
\bibinfo{author}{\bibfnamefont{M.}~\bibnamefont{Martinez}},
  \bibinfo{author}{\bibfnamefont{M.~D.} \bibnamefont{Sievert}},
  \bibinfo{author}{\bibfnamefont{D.~E.} \bibnamefont{Wertepny}},
  \bibnamefont{and}
  \bibinfo{author}{\bibfnamefont{J.}~\bibnamefont{Noronha-Hostler}}
  (\bibinfo{year}{2019}), \eprint{1911.10272}.

\bibitem[{\citenamefont{Churchill et~al.}(2020)\citenamefont{Churchill, Yan,
  Jeon, and Gale}}]{Churchill:2020uvk}
\bibinfo{author}{\bibfnamefont{J.}~\bibnamefont{Churchill}},
  \bibinfo{author}{\bibfnamefont{L.}~\bibnamefont{Yan}},
  \bibinfo{author}{\bibfnamefont{S.}~\bibnamefont{Jeon}}, \bibnamefont{and}
  \bibinfo{author}{\bibfnamefont{C.}~\bibnamefont{Gale}}
  (\bibinfo{year}{2020}), \eprint{2008.02902}.

\bibitem[{\citenamefont{Kasmaei and Strickland}(2019)}]{Kasmaei:2018oag}
\bibinfo{author}{\bibfnamefont{B.~S.} \bibnamefont{Kasmaei}} \bibnamefont{and}
  \bibinfo{author}{\bibfnamefont{M.}~\bibnamefont{Strickland}},
  \bibinfo{journal}{Phys. Rev. D} \textbf{\bibinfo{volume}{99}},
  \bibinfo{pages}{034015} (\bibinfo{year}{2019}), \eprint{1811.07486}.

\bibitem[{\citenamefont{Kasmaei and Strickland}(2020)}]{Kasmaei:2019ofu}
\bibinfo{author}{\bibfnamefont{B.~S.} \bibnamefont{Kasmaei}} \bibnamefont{and}
  \bibinfo{author}{\bibfnamefont{M.}~\bibnamefont{Strickland}},
  \bibinfo{journal}{Phys. Rev. D} \textbf{\bibinfo{volume}{102}},
  \bibinfo{pages}{014037} (\bibinfo{year}{2020}), \eprint{1911.03370}.

\end{thebibliography}
\end{document}